\documentclass[twoside]{report}
\usepackage{iwsm}
\usepackage{graphicx}
\usepackage{amsmath, amssymb}
\usepackage{booktabs}

\usepackage{tikz}
\usetikzlibrary{bayesnet}
\usepackage{bm}

\begin{document}


\title{Handling outcome-dependent missingness with binary responses: A Heckman-like model}
\titlerunning{A Heckman-like model for binary outcomes}

\author{Marco Doretti\inst{1}, Elena Stanghellini\inst{2}, Alessandro Taraborrelli\inst{2}}
\authorrunning{Marco Doretti et al.}    

\institute{University of Florence, Italy
\and University of Perugia, Italy}

\email{marco.doretti@unifi.it}

\abstract{In regression models with missing outcomes, selection bias can arise when the missingness mechanism depends on the outcome itself. This proposal focuses on an extension of the Heckman model to a setting where the outcome is binary and both the selection process and the outcome are modeled through logistic regression. A correction term analogous to the inverse Mills' ratio is derived based on relative risks. Under given assumptions, such a strategy provides an effective tool for bias correction in the presence of informative missingness.}


\maketitle



\section{Introduction}
Non-ignorable missing outcome in a regression framework is quite common in applied settings, and adjusting methods are needed in order to avoid bias within the estimation process. In this framework, the Heckman model (Heckman 1976, 1979) has become a very popular tool to correct for informative missingness of a continuous outcome in a linear model. As it is well-known, the whole strategy relies on specifying a regression model for the missingness indicator, which is subsequently used to build a correction term to be included in the outcome model, fitted on the non-random sub-sample of complete records. Extensions of Heckman's model exist to accommodate a binary response; see for example Van de Ven and Van Praag (1981) where the outcome equation follows a Probit model.

In this work, we investigate the distortion induced by a missingness mechanism that is influenced by the outcome itself, together with other covariates, within a logistic regression framework. In particular, a logistic regression model is postulated for both the outcome and the missingness indicator, with the primary focus of recovering the parameters of the former. An approximated procedure is proposed that works well under the assumption that the outcome in the sub-sample of units for which the outcome is available is either rare or frequent within covariate strata, a fact that can be easily checked. We show that the parameters of the missing equation can be estimated from the marginal model by introducing an additional covariate that is the estimated probability of the outcome in the observed data, whose semi-parametric expression can be derived. Finally, in line with Heckman's idea, we include in the outcome model (fitted on the complete data) a relative risk term obtained from the missingness model parameters.

\section{The model}

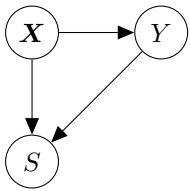
\begin{figure}[tb]
    \centering
    \begin{tikzpicture}
        \node[latent] (X) {$\bm{X}$};
       \node[latent, right=of X] (Y) {$Y$};
        \node[latent, below=of X] (S) {$S$};

        \edge {X} {Y}; 
        \edge {X, Y} {S}; 
    \end{tikzpicture}
    \caption{Data generating process: $Y$, outcome, $S$: missingness indicator, $\bm{X}$: covariates.}
    \label{fig:dag}
\end{figure}

For the $i$-th unit of the sample at hand ($i=1,\dots,n$), we denote by $\bm{X}_i$ the vector of covariates, by $Y_{i}$ the binary outcome that can be either observed or missing, and by $S_i$ an indicator variable that takes value 1 if $Y_i$ is observed and 0 otherwise. The postulated logistic models for the outcome and for the missingness mechanism are respectively given by
\begin{equation}\label{eq:out}
    \textup{logit}\{P(Y_i=1 \mid \bm{X}_i=\bm{x}_i)\} = \beta_0 + \bm{x}_i^\T\bm{\beta}_x
\end{equation}
and
\begin{equation}\label{eq:sel}
    \textup{logit}\{P(S_i=1 \mid \bm{X}_i=\bm{x}_i,Y_i=y_i)\} = \delta_0 + \bm{x}_i^\T\bm{\delta}_x+ \delta_y y_i,
\end{equation}
with the corresponding data generating process depicted in Figure~\ref{fig:dag}.

Clearly, the above models are not directly estimable due to the presence of missing data. However, based on results in Doretti et al. (2024) it is possible to derive the sub-sample outcome model
\begin{equation}\label{eq:outsel}
    \textup{logit}\{P(Y_i=1 \mid \bm{X}_i=\bm{x}_i,S_i=1)\} = 
    \beta_0 +  \bm{x}_i^\T\bm{\beta}_x + \log \textup{RR}_{S_i=1 \mid Y_i,\bm{x}_i},
\end{equation}
where, for two binary variables $A$ and $B$ and a random vector $\bm{C}$,
\[
\textup{RR}_{A=a \mid B,\bm{c}}=\frac{P(A=a\mid B=1,\bm{C}=\bm{c})}{P(A=a\mid B=0,\bm{C}=\bm{c})}
\]
denotes, for every configuration $\bm{C}=c$, the relative risk of the event $[A=a]$ when varying $B$ across its two levels.

In practice, the logarithmic term in~\eqref{eq:outsel} is analogous to the inverse Mills' ratio in Heckman's model, in that it corrects the distortion due to the missingness mechanism via regression adjustment. However, such a term depends on the parameter vector $\bm{\delta}=(\delta_0,\bm{\delta}_x^\T,\delta_y)^\T$, for which an identification strategy beyond the model~\eqref{eq:sel} is needed. To this end, one can again exploit the relationship between marginal and conditional logistic models to write  
\begin{equation}
\label{eq:selmarg}
\begin{split}
         \textup{logit}\{P(S_i=1 \mid \bm{X}_i=\bm{x}_i)\} &= \delta_0 + \bm{x}_i^\T\bm{\delta}_x + \delta_y - \log RR_{Y_i=1 \mid S,\bm{x}} \\
         &= \delta_0 + \bm{x}_i^\T\bm{\delta}_x - \log RR_{Y_i=0 \mid S,\bm{x}}.
\end{split}
\end{equation}

Like in Equation~\eqref{eq:outsel}, the logarithmic terms of~\eqref{eq:selmarg} cannot be directly computed since they depend on the unobservable probabilities $P(Y_i=1\mid S_i=0, \bm{X}_i=\bm{x}_i)$ and $P(Y_i=0\mid S_i=0, \bm{X}_i=\bm{x}_i)$, respectively. However, using the parametric expressions in Stanghellini and Doretti (2019), it can be proven that the following relations hold:
\[
\begin{split}
\log \textup{RR}_{Y_i=1 \mid S_i,\bm{x}_i} &= \log [1+ \{\exp(\delta_y)-1\}P(Y_i=0 \mid S_i=1,\bm{X}_i=\bm{x}_i)] \\
\log \textup{RR}_{Y_i=0 \mid S_i,\bm{x}_i} &= \log [1+ \{\exp(-\delta_y)\}-1\}P(Y_i=1 \mid S_i=1,\bm{X}_i=\bm{x}_i)].
\end{split}
\]
Thus, when either $P(Y_i=1 \mid S_i=1,\bm{X}_i=\bm{x}_i) $ or $P(Y_i=0 \mid S_i=1,\bm{X}_i=\bm{x}_i)$ is small, the well-known approximation $\log(1+x) \approx x$ can be used to obtain
\[
\log \textup{RR}_{Y_i=1 \mid S_i,\bm{x}_i} \approx [\exp{(\delta_y)}-1]P(Y_i=0 \mid S_i=1,\bm{X}_i=\bm{x}_i)
\]
or
\[
\log \textup{RR}_{Y_i=0 \mid S_i,\bm{x}_i} \approx [\exp{(-\delta_y)}-1]P(Y_i=1 \mid S_i=1,\bm{X}_i=\bm{x}_i).
\]

As a consequence, the following strategy can be implemented:
\begin{enumerate}
\item Obtain an initial non-parametric or semi-parametric estimate of $P(Y_i=1 \mid S_i=1,\bm{X}_i=\bm{x}_i)$; 
\item Include this estimate as a covariate in Equation~\eqref{eq:selmarg} in order to approximate the correction term and recover the $\bm{\delta}$ vector;
\item Use $\bm{\delta}$ in order to build the correction term to be included as an offset in Equation~\eqref{eq:outsel} and recover $\bm{\beta}=(\beta_0,\bm{\beta}^\T_x)^\T$.
\end{enumerate}
%
%

\section{Concluding remarks}

The proposed method has the main objective to recover the outcome model parameters $\bm{\beta}$. Indeed, partial identification issues might affect the estimation of $\bm{\delta}$ in practice. Another challenge concerns the estimation of $P(Y_i=y \mid S_i=1,\bm{X}_i=\bm{x}_i) $: if a fully parametric method like logistic regression is used, mis-specification problems are likely to arise since, due to non-collapsibility of odds-ratios, such a model is incompatible with the postulated one in~\eqref{eq:out}. A potential solution to this problem is the use of Generalized Additive Models (GAMs) or other semi-parametric strategies. 

The methodology described here might also be extended to a longitudinal setting, in the spirit of the existing proposals for the linear case. Future research could focus on this extension, enabling a more comprehensive analysis of selection dynamics over time.

\references
\begin{description}

\item[Doretti, M., Genbäck, M., and Stanghellini, E.] (2024). Mediation analysis with case-control sampling: Identification and estimation in the presence of a binary mediator. {\it Biometrical Journal}, {\bf 66(1)}, 2300089.
\item[Heckman, J. J.] (1976). The common structure of statistical models of truncation, sample selection and limited dependent variables and a simple estimator for such models. {\it Annals of Economic and Social Measurement}, {\bf 5(4)}, 475\,--\,492.
\item[Heckman, J. J.] (1979). Sample selection bias as a specification error.
     {\it Econometrica}, {\bf 47(1)}, 153\,--\,161.
\item[Stanghellini, E., Doretti, M.] (2019).
     On marginal and conditional parameters in logistic regression models.
     {\it Biometrika}, {\bf 106(3)}, 732\,--\,739.
\item[Van de Ven, W. and Van Praag, B] (1981).
     The demand for deductibles in private health insurance: A probit model with sample selection.
    {\it Journal of Econometrics}, {\bf 17(2)}, 229\,--\,252.
\end{description}

\end{document}